\documentclass[column,preprintnumbers,amsmath,amssymb]{revtex4}
\preprint{}
\usepackage{graphicx}
\usepackage{dcolumn}
\usepackage{bm}
\usepackage{mathrsfs}
\begin{document}

\title{Enhancing parameter estimation precision in dissipative environment with two-photon driving }
\author{Dong  Xie}
\email{xiedong@mail.ustc.edu.cn}
\affiliation{College of Science, Guilin University of Aerospace Technology, Guilin, Guangxi, P.R. China.}

\author{Chunling Xu}
\email{lingling19880210@126.com}
\affiliation{College of Science, Guilin University of Aerospace Technology, Guilin, Guangxi, P.R. China.}

\begin{abstract}
We investigate the frequency estimation of an optical field suffering from an unavoidable dissipative environment. Generally, dissipative noises greatly reduce the precision. Here, we find that two-photon driving can improve the measurement precision by resisting the noises. Moreover, in long time, the uncertainty of frequency can be close to 0 with a proper magnitude of the parametric two-photon drive, which is in sharp contrast to the uncertainty going to infinity without the two-photon driving. Our results show that two-photon driving can realize the ultrasensitive measurement in dissipative environment under the long-encoding-time condition.
\end{abstract}

\maketitle

\section{Introduction}
Improving the precision of parameter measurement plays an unparalleled role in the development of basic science and technology\cite{lab1,lab2,lab3,lab4,lab5,lab6}. In classical physics, the best metrology precision, known as the quantum shot noise limit (SNL), scales as $1/\sqrt{N}$ with $N$ being the number of resource employed in the measurements. Quantum effects can help us to beat the SNL, such as squeezing\cite{lab7,lab8,lab9,lab10,lab11} and entanglement\cite{lab12,lab13,lab14,lab15}.
In particular, the quantum effect of light can enhance the imaging resolution\cite{lab16} in many important fields, such as, radar\cite{lab17,lab18}and gravitational wave detection\cite{lab19,lab20}.

Although quantum effects can improve measurement accuracy, they are actually quite fragile. The main reason is the inevitable environmental interference.
A realistic physical system inevitably interacts with the surrounding environment, leading to lose information from the system to the environment. When the decoherence induced by the environment is severe, the measurement accuracy should not exceed or even far below the SNL. A lot of works have shown that for noisy
quantum metrology\cite{lab21,lab22,lab23,lab24}, initial quantum entangled or squeezed state can not help to obtain better precision than the SNL.
In order to prevent environmental interference, some methods have been proposed.  Dynamical decoupling is an active and effective method\cite{lab25,lab26}, and it has been applied to suppress the decoherence in optical fibers\cite{lab27,lab28} as well as in superconducting Qubits\cite{lab29}.
In general, the strategy of dynamic decoupling has been studied primarily in the $\delta$-pulse regime\cite{lab30,lab31}. Namely, in dynamic decoupling, unitary control pulses are instantaneously applied to the system at specific times. However, while decoherence prevents environmental interference, it also prevents the encoding of parameter information. So dynamic decoupling can not perform well in quantum metrology. It was really found that the non-Markovian effect\cite{lab32} can make the precision surpass the SNL in dephasing noises with Ohmic spectral density. However, as encoding time increases, the precision gets worse and worse. Correlated environments\cite{lab33} can also make the precision surpass the SNL. But, the symmetrical correlated environment is few in reality.

In this article, we use a two-photon driving to improve the frequency precision of the optical field in a general unavoidable dissipative environment. Focusing on the long-encoding-time condition, in the phenomenological description\cite{lab34,lab35} the photons completely lose, leading to that the information of frequency can not be obtained.  We investigate the function of different magnitudes of the parametric drive on frequency estimation.  Our analysis reveals that the uncertainty of frequency can be close to 0 with a proper magnitude of the parametric drive in the long-encoding-time condition.  Our result shows that two-photon driving can realize the ultrasensitive measurement in the dissipative environment by forming the effective non-Hermitian parity-time (PT) symmetry dynamics\cite{lab36,lab37,lab38,lab39,lab40,lab41}.

The article is organized as follows. In Section II, we introduce the physical model and  the mathematical description of two measurement ways. In Section A, we consider the case of small magnitude of the parametric drive in the long-encoding-time condition. In Section B, the case of large magnitude is discussed with the initial coherent state. The case of specific magnitude is discussed in section C. In Section III, a simple explanation is given. We make a brief conclusion and outlook in Section IV. In Appendix A, we give the detail analytical derivation.
\section{frequency measurement in  dissipative noises}
In general, the process of estimating the parameter of a system can be classified as three steps: first step, prepare a initial probe state; second step, encode the parameter information; third step, make the final measurement to obtain the parameter.

Here, we want to measure the frequency $\omega$ of a optical field.  The process is shown in Fig.1. We prepare the photons with the initial pure state $|\psi_{in}\rangle$. Then, the information of the frequency $\omega$ is encoded. In the ideal metrology, the measurement uncertainty scales as $1/\sqrt{T}$, where $T$ denotes the interrogation time. Hence, for the long-encoding-time, one can obtain $\omega$ with high accuracy. However, during the encoding, dissipative environment is unavoidable\cite{lab42,lab43,lab44}. This leads to greater uncertainty over time. The precision of frequency $\omega$ gets worse and worse. We will utilize two-photon driving to improve the result. At the last step, the final detection obtains the information of $\omega$. In Fig.1, we consider that there are two feasible detection schemes: (a) represents direct photon detection, (b) represents homodyne detection. Maybe both of the two detection schemes are not optimal, which can not saturate the Cram$\acute{e}$r-Rao bound governed by quantum Fisher information.  But it sufficiently demonstrates the superiority of a two-photon drive especially in an experimentally friendly manner.
\begin{figure}[h]
\includegraphics[scale=0.4]{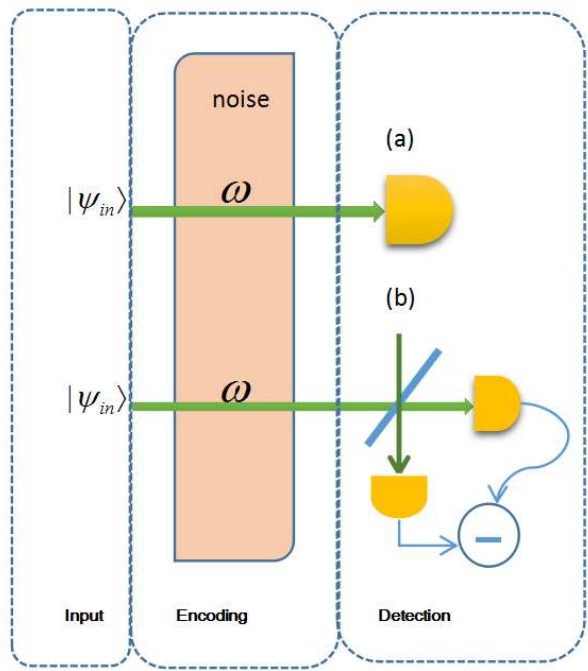}
 \caption{\label{fig.1}Diagram of estimation frequency $\omega$. $|\psi_{in}\rangle$ represents the initial state.  (a) represents the direct photon detection, (b) represents the homodyne detection. Dissipative noise has an impact during the process of encoding parameter $\omega$.}
 \end{figure}

The total Hamiltonian is described by ($\hbar=1$ throughout this article )

\begin{eqnarray}
\widehat{H}=\omega \hat{a}^\dagger \hat{a}+\sum_k[\omega_k\hat{b}_k^\dagger \hat{b}_k+g_k(\hat{a}\hat{b}_k^\dagger+\hat{a}^\dagger\hat{b}_k)]+i\frac{\lambda}{2}(\hat{a}^{\dagger2}- \hat{a}^2),
\end{eqnarray}
where $\hat{a}$ is the annihilation operator of the optical field with frequency $\omega$, $\hat{b}_k$ is the annihilation operator of the $k$th environmental mode with frequency $\omega_k$ and  $g_k$ denotes its coupling strength to the optical field. $\lambda$ is the magnitude
of the two-photon drive, which can be realized in  down-conversion processes in nonlinear optics.
The environment structure can be characterized by the spectral density $J(\omega')=\sum_kg_k^2\delta(\omega'-\omega_k)$\cite{lab45}.

In the Heisenberg picture, we can obtain $\hat{a}(t)=G(t)\hat{a}(0)+L^*(t)\hat{a}^\dagger(0)+\sum_k[\mu_k(t)\hat{b}(0)+\nu_k^*(t)\hat{b}^\dagger(0)]$, with $|G(t)|^2+|L(t)|^2+\sum_k(|\mu_k(t)|^2+|\nu_k|^2)=1$. At the same time, $G(t)$, $L(t)$, $\mu_k(t)$ and $\nu_k(t)$ must satisfy following equations

\begin{eqnarray}
\dot{G}(t)&=&\lambda L(t)-i\omega G(t)-\int_0^\infty ds K(t-s)G(s),\\
\dot{L}(t)&=&\lambda G(t)+i\omega L(t)-\int_0^\infty ds K^*(t-s)L(s),\\
\dot{\mu}_k(t)&=&\lambda \nu_k-i\omega \mu_k(t)-\int_0^\infty ds K(t-s)\mu_k(s)-ig_ke^{-i\omega t},\\
\dot{\nu}_k(t)&=&\lambda \mu_k-i\omega \nu_k(t)-\int_0^\infty ds K^*(t-s)\nu_k(s),
\end{eqnarray}
where $K(t-s)=\int_0^\infty d\omega' J(\omega')e^{-i\omega'(t-s)}$ is the noise correlation function and the initial conditions $G(0)=1$, $L(0)=0$, $\mu_k(0)=0$, and $\nu_k(0)=0$(see Appendix A).
We suppose that the bandwidth of the interaction spectrum is much larger than the coupling strength. Therefore the Wigner-Weisskopf approximation\cite{lab46,lab47} can be used,  which has been proved to be equivalent to the Markov approximation\cite{lab48}.
By replacing $J(\omega')$ with $J(\omega)$ and extending the lower limit of $¦Ø$ in the integral to be $-\infty$\cite{lab46,lab47}, we can obtain

\begin{eqnarray}
K(t-s)=2\gamma\delta(t-s),
\end{eqnarray}
where $\gamma=\pi J(\omega)$.
Then we can obtain the analytical results of  $G(t)$, $L(t)$, $\mu_k(t)$, and $\nu_k(t)$  by Laplace transform

\begin{eqnarray}
G(t)=e^{-\gamma t}(\cosh[t\sqrt{\lambda^2-\omega^2}]-\frac{i\omega\sinh[t\sqrt{\lambda^2-\omega^2}]}{\sqrt{\lambda^2-\omega^2}}),\
L(t)=\frac{\lambda e^{-\gamma t}\omega\sinh[t\sqrt{\lambda^2-\omega^2}]}{\sqrt{\lambda^2-\omega^2}},
\end{eqnarray}
The detail analytical results of $\mu_k(t)$ and $\nu_k(t)$ are shown in Appendix A.

Then the best precision of estimating $\omega$ can be evaluated by the error propagation formula

\begin{eqnarray}
&\delta\omega=\frac{\delta M}{|\partial \overline{M}/\partial\omega|},
\end{eqnarray}
in which
\begin{eqnarray}
\delta M=\sqrt{\langle e^{iH t}M^2e^{-iH t}\rangle-\overline{M}^2},\    \overline{M}=\langle e^{iH t}Me^{-iH t}\rangle,
\end{eqnarray}
where $\langle\bullet\rangle=\langle \Psi_E(0)|\langle\psi_{in}|\bullet|\psi_{in}\rangle|\Psi_E(0)\rangle$. We assume that the noise is initially in vacuum state $|\Psi_E(0)\rangle=|\{0_k\}\rangle$.
For direct photon detection in Fig.1, the measurement operator $M_d=\hat{a}^\dagger \hat{a}$; for homodyne detection, the measurement operator is the field quadrature $M_h=\frac{e^{-i\theta}\hat{a}^\dagger+ \hat{a}e^{i\theta}}{2}$, where measurement angle $\theta$ can be controlled by the local oscillator\cite{lab499}.
\subsection{Small magnitude of the parametric drive in the long-encoding-time condition}
In the ideal metrology, the uncertainty of parameter $\omega$ is proportional to $1/t$, where $t$ denotes the interrogation time.
The measurement precision of $\omega$ is close to 0  for the long-encoding-time. However, in the Markovian noise, the information of parameter  disappears completely for long time\cite{lab49}. In this section, we consider that the precision is obtained with a small magnitude $\lambda$ of the parametric drive in the long-encoding-time condition. When $\lambda^2<\gamma^2+\omega^2$, we define it as the small magnitude.

For the long-encoding-time, $$t[\gamma-(\lambda^2-\omega^2)^{1/2}]\gg1,$$ we obtain that
\begin{eqnarray}
\hat{a}(t)=\sum_k\frac{g_k(-i \gamma - \omega - \omega_k)e^{-i\omega_kt}}{\lambda^2 - \omega^2 - (r - i \omega_k)^2}\hat{b}_k+\frac{g_k\lambda e^{i\omega_kt}}{\lambda^2 - \omega^2 - (\gamma + i \omega_k)^2}\hat{b}^\dagger_k,
\end{eqnarray}

The information of $\omega$ can be obtained by the direct photon detection, but not by the homodyne detection due to $\overline{M_h}=0$.
With the direct photon detection, we can derive that (see Appendix B)
\begin{eqnarray}
&\overline{{M_d}}&=\langle \hat{a}^\dagger(t)\hat{a}(t)\rangle=\frac{\lambda^2}{-\lambda^2+\omega^2+\gamma^2},\\
&\delta M_d&\approx\frac{\lambda^2(3\gamma^2+3\omega^2-\lambda^2)}{(-\lambda^2+\omega^2+\gamma^2)^2}.
\end{eqnarray}
Substituting above equations into Eq.(10), the uncertainty of $\omega$ is derived
\begin{eqnarray}
\delta\omega^2\approx\frac{(-\lambda^2 + \gamma^2 +\omega^2)^2 [-\lambda^2 +3 (\gamma^2 +\omega^2)]}{4 \lambda^2 w^2}.
\end{eqnarray}
We can see that without the two-photon drive($\lambda=0$) the uncertainty $\delta\omega=\infty$, which means that the information of parameter disappears completely. With the two-photon drive($\gamma\neq0$) the uncertainty is a finite value, which means that we can still obtain the information of parameter. In particular, when $\lambda^2\rightarrow\gamma^2+\omega^2$,  the uncertainty $\delta\omega\approx0$. It shows that the parameter $\omega$ can be measured accurately. And the estimation precision is independent of the initial state of the optical field.
\subsection{Large magnitude of the parametric drive}
In this section, we investigate the estimation precision with large magnitude, $\lambda^2>\gamma^2+\omega^2$.
For the long-encoding-time, $t[(\lambda^2-\omega^2)^{1/2}-\gamma]\gg1$, we obtain that
\begin{eqnarray}
\hat{a}(t)=e^{(\sqrt{\lambda^2-\omega^2}-\gamma)t}\{\frac{1}{2}(1-i\frac{\omega}{\sqrt{\lambda^2-\omega^2}})\hat{a}+\frac{\lambda}{2\sqrt{\lambda^2-\omega^2}}\hat{a}^\dagger+\sum_k[\frac{- g_k [i \lambda^2 + (-i \omega + \sqrt{\lambda^2 - \omega^2}) (i \gamma + \omega + \omega_k)]}{2 \sqrt{\lambda^2-\omega^2} [\lambda^2 - \omega^2 - (\gamma +i \omega_k)^2]}\hat{b}_k\nonumber\\
+\frac{i \lambda g_k (\gamma + \sqrt{\lambda^2 - \omega^2}+
   i \omega_k)}{2 \sqrt{\lambda^2-\omega^2} [\lambda^2 - \omega^2 - (\gamma +i \omega_k)^2]}\hat{b}^\dagger_k]\},
\end{eqnarray}

When the initial state is coherent state $|\psi_{in}\rangle=|\alpha\rangle$, $\alpha^2\gg1$, we use two ways to measure the parameter $\omega$.
For the direct photon detection, the uncertainty is described by (see Appendix B)
\begin{eqnarray}
\delta\omega^2\approx\frac{\lambda(\lambda - \omega) (\lambda + w)}{4 Nt^2 w^2 (\lambda +\sqrt{\lambda^2 - \omega^2})},
\end{eqnarray}
where $N=\alpha^2$ denotes the number of input photons. We can see that $\delta\omega^2\propto 1/(Nt^2)$, recovering the scale in the ideal metrology. And from above equation, we can see that the uncertainty of the magnitude $\lambda$ will increase with the magnitude $\lambda$. This shows that the optimal magnitude should be close to $\sqrt{\gamma^2+\omega^2}$.

From Eq.(15), the direct photon detection obtain low precision for small value of $\omega$, especially for $\omega\simeq0$. In order to solve this question, we consider the homodyne detection. By the similar calculation, we obtain the uncertainty of $\gamma$ with the homodyne detection
\begin{eqnarray}
\delta\omega^2\approx\frac{\lambda^2}{16Nt^2\omega^2},\ \textmd{for}\ \theta=0,\\
\delta\omega^2\approx\frac{\lambda^2}{8Nt^2},\ \textmd{for}\ \theta=\pi/2,
\end{eqnarray}
where we considering $\lambda\gg\omega>0$ and $\omega^2t\gg\lambda^2$. This result shows that the homodyne detection with the angle $\theta=0$ performs like the case of the direct photon detection. Homodyne detection with the angle $\theta=\pi/2$ can perform very well in low value of parameter $\omega$.
It's worth mentioning that for $\omega\simeq0$, the estimation precision is given by a similar calculation
\begin{eqnarray}
\delta\omega^2\approx0, \ \textmd{for}\ \theta=\pi/2.
\end{eqnarray}

\subsection{Specific magnitude of the parametric drive}
Above two sections show that the closer  magnitude $\lambda$ gets to $\sqrt{\gamma^2+\omega^2}$, the higher the accuracy is. But the premise is that $\lambda$ is not equal to $\sqrt{\gamma^2+\omega^2}$ in the above two sections.
In this section, we discuss about the case of $\lambda=\sqrt{\gamma^2+\omega^2}$.

For long time $t\gamma\gg1$, the corresponding annihilation operator of optical field in the Heisenberg picture  is described by
\begin{eqnarray}
\hat{a}(t)=\frac{1}{2}(1-i\frac{\omega}{\sqrt{\lambda^2-\omega^2}})\hat{a}+\frac{\lambda}{2\sqrt{\lambda^2-\omega^2}}\hat{a}^\dagger+\sum_k\{\frac{- g_k [i \lambda^2 + (-i \omega + (1-2e^{-i2\omega_kt})\sqrt{\lambda^2 - \omega^2}) (i \gamma + \omega + \omega_k)]}{2 \sqrt{\lambda^2-\omega^2} [\lambda^2 - \omega^2 - (\gamma +i \omega_k)^2]}\hat{b}_k
\\+\frac{i \lambda g_k (-2e^{i2\omega_kt}\sqrt{\lambda^2 - \omega^2}+\gamma + \sqrt{\lambda^2 - \omega^2}+
   i \omega_k)}{2 \sqrt{\lambda^2-\omega^2} [\lambda^2 - \omega^2 - (\gamma +i \omega_k)^2]}\hat{b}^\dagger_k\},
\end{eqnarray}

By using the initial coherent state $|\psi_{in}\rangle=|\alpha\rangle$ and performing the similar calculation in Section B, we can obtain the estimation precision for $\gamma t\gg1$ and $\omega>0$ (see Appendix B)
\begin{eqnarray}
\delta\omega^2\approx\frac{\gamma^2 (7 \gamma^2 + 3 \omega^2)}{4 Nt^2 \omega^2 (\gamma + \sqrt{\gamma^2 + \omega^2})^2},\ \textmd{for}\ \theta=0\\
\delta\omega^2\approx(7 \gamma^4 + 3 \gamma^2 \omega^2)/(4N t^2 \omega^4),\ \textmd{for}\ \theta=\pi/2.
\end{eqnarray}
For $\omega=0$, the estimation can be given by $\delta\omega^2\approx7\gamma^2/(4N)$ with the measurement angle $\theta=\pi/2$.

Comparing with the results in Section A and Section B, we find that only for small value of $\gamma$ ($\gamma\ll\omega$), the estimation with the magnitude $\lambda=\sqrt{\gamma^2+\omega^2}\approx\omega$ is optimal for estimating the frequency $\omega$. With the homodyne detection, the measurement precision obtained at the specific magnitude $\lambda=\sqrt{\gamma^2+\omega^2}$ is slightly lower than that obtained near the specific magnitude $\lambda=\sqrt{\gamma^2+\omega^2}$ for large value $\gamma$. For any value of $\gamma$, it is worth further exploring whether the optimal measurement precision can be obtained at point $\lambda=\sqrt{\gamma^2+\omega^2}$ by optimizing the measurement scheme (we leave it as an open question).

To sum up, with the two-photon drive ($\lambda>0$), the uncertainty of frequency is becoming finite for long time, which  is in stark contrast to the results without a drive. When $\gamma$ is close to or equal to $\sqrt{\gamma^2+\omega^2}$, the uncertainty of frequency $\omega$ can be close to 0 for $\omega\neq0$ and long time $t$. This means that direct photon detection and homodyne detection approach the optimal measurement scheme under the long-encoding-time condition and $\omega\neq0$.

\section{discussion}
We have investigated the estimation precision of the frequency $\omega$ for the long encoding time. The results show that the two-photon drive can help to improve the estimation. When  $\gamma\ll\omega$, the optimal magnitude is at the specific point $\lambda=\sqrt{\gamma^2+\omega^2}\approx\omega$. Otherwise the magnitude $\lambda=\omega$ is not optimal in estimating the frequency $\omega$.
We can give an explanation with the PT symmetry dynamics. Defining a vector of operators
\begin{eqnarray}
|\hat{a}\rangle=(\hat{a},\hat{a}^\dagger)^T
\end{eqnarray}
the Heisenberg equations of motions can be written as
\begin{eqnarray}
i\partial_t|\hat{a}\rangle=\hat{H}_{\textmd{eff}}|\hat{a}\rangle+(\hat{F}, \hat{F}^\dagger)^T,
\end{eqnarray}
where $\hat{F}=\sum_kg_k\hat{b}_k(t)$ and the effective Hamiltonian  $\hat{H}_{eff}$ is
\[
 \hat{H}_{\textmd{eff}}= \left(
\begin{array}{ll}
\ \omega\ \ \ \ \ i \lambda\\
i \lambda\ \ \   -\omega
  \end{array}
\right ).
\]
This effective Hamiltonian is non-Hermitian PT symmetrical.  The eigenvalues are written  as $\pm\sqrt{\omega^2-\lambda^2}$. The exceptional point is at $\omega=\lambda$\cite{lab500}.
A lot of works \cite{lab36,lab37,lab38,lab39,lab40,lab41} have shown that the exceptional point can improve the estimation precision.
So for $\gamma\ll\omega$, in Section C, we prove that the optimal magnitude is $\lambda=\sqrt{\omega^2+\gamma^2}\approx\omega$.
However, when $\sqrt{\omega^2+\gamma^2}$ is not close to $\omega$, the magnitude $\lambda=\omega$  is not optimal.   It is because that quantum-limited signal to noise at EPs is proportional to the perturbation from the environment\cite{lab501}.

According to the calculations in Section A-C, when the magnitude $\lambda$ is near $\sqrt{\omega^2+\gamma^2}$, the uncertainty is close to 0. The reason is that when the magnitude is close to $\sqrt{\omega^2+\gamma^2}$, the number of exciting photons $\langle \hat{a}^\dagger(t)\hat{a}(t)\rangle$ is close to infinity.
So considering the cost of energy (or time), the optimal estimation precision of $\omega$ should not be equal to 0 exactly.

\section{conclusion and outlook}
We have investigated the function of the two-photon drive on improving the estimation precision of field frequency. By applying the Wigner-Weisskopf approximation, we obtain the analytical results of dynamics of field operator in Heisenberg picture. With the direct photon detection and the homodyne detection, we reveal that devastating results from the dissipation noise can be suppressed. With the two-photon drive the uncertainty of frequency is becoming finite for long time, which  is in stark contrast to the results without a drive. Moreover, with long encoding time the estimation uncertainty can be close to 0 for the magnitude close to $\sqrt{\omega^2+\gamma^2}$.

The present study is expected to impact deeper understanding the two-photon driving. It is helpful for high precision sensor design. Although our research is in quantum optical systems, it can be extended to general Bose systems that are parametrically driven. The upcoming work is on the role of quantum resources, such as quantum entangled or squeezed state, in improving the accuracy of parameter measurements in noisy environments using two-photon drives. At the same time, the role of two-photon drive in non-Markov environments is also worth investigating.

\section*{Acknowledgement}
 This research was supported by the National
Natural Science Foundation of China under Grant No. 11747008 and Guangxi Natural Science Foundation 2016GXNSFBA380227.

\section*{Appendix A:}
We give the detailed derivation of the dynamical equation under the total Hamiltonian
\begin{eqnarray}
\widehat{H}=\omega \hat{a}^\dagger \hat{a}+\sum_k[\omega_k\hat{b}_k^\dagger \hat{b}_k+g_k(\hat{a}\hat{b}_k^\dagger+\hat{a}^\dagger\hat{b}_k)]+i\frac{\lambda}{2}(\hat{a}^{\dagger2}- \hat{a}^2).
\end{eqnarray}

In the Heisenberg picture, the equations of motion of the field operator read

\begin{eqnarray}
\dot{\hat{a}}(t)&=&\lambda\hat{a}^\dagger(t)-i\omega\hat{a}(t)-i\sum_kg_k\hat{b}_k(t),\\
\dot{\hat{b}}_k(t)&=&-i\omega_k\hat{b}_k(t)-ig_k\hat{a}(t).
\end{eqnarray}
Substituting the formal solution of Eq.(27)
\begin{eqnarray}
\dot{\hat{b}}_k(t)=\hat{b}_k(0)e^{-\omega_kt}-ig_k\int_0^te^{-\omega_k(t-s)}\hat{a}(s)ds
\end{eqnarray}
into Eq.(26), we obtain
\begin{eqnarray}
\dot{\hat{a}}(t)=\lambda\hat{a}^\dagger(t)-i\omega\hat{a}(t)-\int_0^tdsK(t-s)\hat{a}(s)-i\sum_kg_k\hat{b}_k(0)e^{-i\omega_kt},
\end{eqnarray}
where $K(t-s)=\int_0^\infty d\omega' J(\omega')e^{-i\omega'(t-s)}$ is the noise correlation function.  Using the Wigner-Weisskopf approximation $K(t-s)=\pi J(\omega)\delta(t-s)$.
The linearity of Eq. (24) implies that
a general field operator $\hat{a}(t)$ can be expanded as
\begin{eqnarray}
\hat{a}(t)=G(t)\hat{a}(0)+L^*(t)\hat{a}^\dagger(0)+\sum_k\mu_k(t)\hat{b}_k(0)+\nu_k^*(t)\hat{b}_k^\dagger(0).
\end{eqnarray}
We  can obtain $|G(t)|^2+|L(t)|^2+\sum_k(|\mu_k(t)|^2+|\nu_k(t)|^2)=1$ from the commutation relation $[\hat{a}(t),\hat{a}^\dagger(t)]=1$.
Substituting this expansion into Eq.(29), the Eqs.(2-5) in Section II can be derived.

By Laplace transform, the Eqs.(2-5) can be solved analytically
\begin{eqnarray}
&&G(t)=e^{-\gamma t}(\cosh[t\sqrt{\lambda^2-\omega^2}]-\frac{i\omega\sinh[t\sqrt{\lambda^2-\omega^2}]}{\sqrt{\lambda^2-\omega^2}},\
L(t)=\frac{\lambda e^{-\gamma t}\omega\sinh[t\sqrt{\lambda^2-\omega^2}]}{\sqrt{\lambda^2-\omega^2}}),\\
&&\mu_k(t)=g_ke^{-t (r + \sqrt{\lambda^2-\omega^2})}\ast \nonumber\\
&&\frac{  \{-i \omega^2 ( e^{2 t \sqrt{\lambda^2 -\omega^2}}-1) [i ( e^{2 t \sqrt{\lambda^2 - \omega^2}}-1) \omega - (1 + e^{
         2 t \sqrt{\lambda^2 - \omega^2}} -2 e^{t (r + \sqrt{a^2 - w^2} - i \omega_k)}) \sqrt{\lambda^2 - \omega^2}] (i r +
       \omega+ w_k)\}}{2 \sqrt{\lambda^2- \omega^2} [\lambda^2 - \omega^2 - (\gamma - i \omega_k)^2]},\\
&&\nu_k(t)=-i \lambda g_k e^{-t (r + \sqrt{\lambda^2-\omega^2})}\ast \nonumber\\
&&\frac{ \{\sqrt{\lambda^2-\omega^2}+ e^{2 t \sqrt{\lambda^2 - \omega^2}} \sqrt{\lambda^2 - \omega^2} -
    2 e^{t (\gamma + \sqrt{\lambda^2 - \omega^2} -i \omega_k)}
      \sqrt{\lambda^2 - \omega^2} + (-1 + e^{2 t \sqrt{\lambda^2 - \omega^2}}) (\gamma -
       i w_k)\}}{2 \sqrt{\lambda^2 - \omega^2} [\lambda^2 - \omega^2- (\gamma -i \omega_k)^2]}.
\end{eqnarray}
\section*{Appendix B:}
For small magnitude and long time, the field operator can be simplified as
\begin{eqnarray}
\hat{a}(t)=\sum_k\frac{g_k(-i \gamma - \omega - \omega_k)e^{-i\omega_kt}}{\lambda^2 - \omega^2 - (r - i \omega_k)^2}\hat{b}_k+\frac{g_k\lambda e^{i\omega_kt}}{\lambda^2 - \omega^2 - (\gamma + i \omega_k)^2}\hat{b}^\dagger_k,
\end{eqnarray}
The  variance of direct photon detection $M_d$ can be calculated as
\begin{eqnarray}
\delta M_d^2=\langle (\hat{a}^\dagger(t)\hat{a}(t))^2\rangle-(\langle \hat{a}^\dagger(t)\hat{a}(t)\rangle)^2,
\end{eqnarray}
where $\langle\bullet\rangle=\langle \Psi_E(0)|\langle\psi_{in}|\bullet|\psi_{in}\rangle|\Psi_E(0)\rangle$. In order to calculate the above equation, we use the
decoupling relation\cite{lab50},  which is written as

\begin{eqnarray}
\langle \hat{A}\hat{B}\hat{C}\hat{D}\rangle\approx\langle\hat{A}\hat{B}\rangle\langle \hat{C}\hat{D}\rangle+\langle\hat{A}\hat{D}\rangle\langle \hat{B}\hat{C}\rangle+\langle\hat{A}\hat{C}\rangle\langle \hat{B}\hat{D}\rangle-2\langle\hat{A}\rangle\langle\hat{B}\rangle\langle \hat{C}\rangle\langle\hat{D}\rangle.
\end{eqnarray}

 Using the Wigner-Weisskopf approximation, we can obtain
\begin{eqnarray}
\sum_k|\frac{g_k(-i \gamma - \omega - \omega_k)e^{-i\omega_kt}}{\lambda^2 - \omega^2 - (r - i \omega_k)^2}|^2=\int_{-\infty}^\infty d\omega_kJ(\omega)|\frac{(-i \gamma - \omega - \omega_k)}{\lambda^2 - \omega^2 - (r - i \omega_k)^2}|^2=1+\frac{\lambda^2}{2(\gamma^2+\omega^2-\lambda^2)},\\
\sum_k|\frac{g_k\lambda e^{i\omega_kt}}{\lambda^2 - \omega^2 - (\gamma + i \omega_k)^2}|^2=\int_{-\infty}^\infty d\omega_kJ(\omega)|\frac{\lambda }{\lambda^2 - \omega^2 - (\gamma + i \omega_k)^2}|^2=\frac{\lambda^2}{2(\gamma^2+\omega^2-\lambda^2)},\\
\sum_k\frac{g_k(-i \gamma - \omega - \omega_k)e^{-i\omega_kt}}{\lambda^2 - \omega^2 - (r - i \omega_k)^2}\frac{g_k\lambda e^{-i\omega_kt}}{\lambda^2 - \omega^2 - (r + i \omega_k)^2}=\frac{\lambda(\gamma+i\omega)}{2(\gamma^2+\omega^2-\lambda^2)}.
\end{eqnarray}

Using above equations we can obtain the uncertainty of parameter $\omega$
\begin{eqnarray}
\delta\omega^2\approx\frac{(-\lambda^2 + \gamma^2 +\omega^2)^2 [-\lambda^2 +3 (\gamma^2 +\omega^2)]}{4 \lambda^2 w^2}.
\end{eqnarray}

Utilizing the similar ways and $\hat{a}|\psi_{in}\rangle=\alpha|\alpha\rangle$ ($\alpha$ is real), we can obtain the results for large magnitude $\lambda>\sqrt{\omega^2+\gamma^2}$.
\textmd{With direct photon detection, the estimation precision is given by}
\begin{eqnarray}
\delta\omega^2\approx\frac{\lambda (\lambda^2 -\omega^2)^3 (\lambda + \sqrt{
  \lambda^2 - \omega^2})}{N\omega^2 [-2 \lambda^3 t + 2 \lambda (t\omega^2 + \sqrt{\lambda^2 - \omega^2}) +
   \lambda^2 (1 - 2 t \sqrt{\lambda^2 -\omega^2}) +\omega^2 (-1 + 2 t \sqrt{\lambda^2 - \omega^2})]^2},
\end{eqnarray}
where $N=\alpha^2$ denotes the number of input photons.

\textmd{With the homodyne detection, the estimation precision is given by}
\begin{eqnarray}
\delta\omega^2\approx\frac{\lambda  (\lambda - \omega)^2 (\lambda +
   \omega)^2 [(\lambda^2 - \omega^2 - 2 \gamma \sqrt{\lambda^2 - \omega^2}) \cos{
     2 \theta} + (-2 \gamma + \sqrt{\lambda^2 - \omega^2}) (\lambda +
      \omega\sin[2 \theta])]}{2N (-\gamma+
   \sqrt{\lambda^2 - \omega^2}) [\omega (\lambda - 2\lambda^2 t + 2 t \omega^2 -
      2 \lambda t \sqrt{\lambda^2 - \omega^2}) \cos{\theta} + (\lambda^2 -
      2 t \omega^2 \sqrt{\lambda^2 - \omega^2}) \sin{\theta}]^2}.
\end{eqnarray}
At the specific magnitude $\lambda=\sqrt{\omega^2+\gamma^2}$, using the similar calculation can derive 
\begin{eqnarray}
\delta\omega^2\approx\frac{ \gamma^4 (7 \gamma^2 + 3\omega^2 +
   2\omega \sqrt{\gamma^2 + \omega^2}
      \sin{
     2 \theta})}{4 N[\omega (\gamma^2 t + (-1 + \gamma t) \sqrt{
       \gamma^2 +
       \omega^2}) \cos\theta - (\gamma^2 + (1 - \gamma t) \omega^2) \sin\theta]^2}
\end{eqnarray}
Under different specific conditions, above equations can  be further simplified in section B and C.


\begin{thebibliography}{9}

\vspace{3mm}
\bibitem{lab1}Giovanetti V., Lloyd S., Maccone L.: Quantum-enhanced measurements: beating the standard quantum limit Science. 306, 1330  (2004)
\bibitem{lab2}Giovannetti V., Lloyd, S., and Maccone L.: Quantum Metrology. Phys. Rev. Lett. 96, 010401 (2006)
\bibitem{lab3}Bongs, K., Launay, R., Kasevich, M. A.: High-order inertial phase shifts for time-domain atom interferometers. Appl. Phys. B 84, 599 (2006)
\bibitem{lab4}Paris, M. G. A.: Quantum Estimation for Quantum Technology. Int. J. Quant. Inf. 7, 125 (2009)
\bibitem{lab5}Taylor, M. A., Janousek, J., Daria, V., Knittel, J., Hage, B., Bachor, H.-A., Bowen, W. P.: Biological measurement beyond the quantum limit. Nat. Photon. 7, 229 (2013)
\bibitem{lab6}S. Slussarenko, M. M. Weston, H. M. Chrzanowski, L. K. Shalm, V. B. Verma, S. W. Nam, and G. J. Pryde, Nat. Photonics 11, 700 (2017).
\bibitem{lab7}M. Xiao, L. A. Wu, and H. J. Kimble, Precision Measurement Beyond the Shot-Noise Limit, Phys. Rev. Lett. 59, 278
(1987).
\bibitem{lab8}P. Grangier, R. E. Slusher, B. Yurke, and A. LaPorta, Squeezed-Light-Enhanced Polarization Interferometer, Phys. Rev. Lett. 59, 2153 (1987).
\bibitem{lab9}K. Lange, J. Peise, B. L$\ddot{u}$cke, L. Pezz$\grave{e}$, J. Arlt, W. Ertmer, C. Lisdat, L. Santos, A. Smerzi, and C. Klempt, Improvement of an Atomic Clock Using Squeezed Vacuum, Phys. Rev. Lett. 117, 143004 (2016).
\bibitem{lab10}R. Schnabel, N. Mavalvala, D. E. McClelland, and P. K. Lam, Quantum metrology for gravitational wave astronomy,
Nat. Commun. 1, 121 (2010).
\bibitem{lab11}J. Aasi et al., Enhanced sensitivity of the LIGO gravitational wave detector by using squeezed states of light, Nat. Photonics 7, 613 (2013).
\bibitem{lab12}P. Cappellaro, J. Emerson, N. Boulant, C. Ramanathan, S. Lloyd, and D. G. Cory, Phys. Rev. Lett. 94, 020502 (2005)
\bibitem{lab13}T. Nagata, R. Okamoto, J. L. O¡¯Brien, K. Sasaki, and
S. Takeuchi, Science 316, 726 (2007).
\bibitem{lab14}Y. Israel, S. Rosen, and Y. Silberberg, Phys. Rev. Lett. 112, 103604 (2014).
\bibitem{lab15}X.-Y. Luo, Y.-Q. Zou, L.-N. Wu, Q. Liu, M.-F. Han, M. K. Tey, and L. You, Science 355, 620 (2017).
\bibitem{lab16}Mankei Tsang, Conservative classical and quantum resolution limits for incoherent imaging, Journal of Modern Optics, 65:11, 1385-1391  (2018).
\bibitem{lab17}S. Barzanjeh, S. Guha, C. Weedbrook, D. Vitali, J. H. Shapiro, and S. Pirandola, Phys. Rev. Lett. 114, 080503 (2015).
\bibitem{lab18}C. W. Sandbo Chang, A. M. Vadiraj, J. Bourassa, B. Balaji, and C. M. Wilson, Appl. Phys. Lett. 114 112601 (2019).
\bibitem{lab19}H. Grote, K. Danzmann, K. L. Dooley, R. Schnabel, J. Slutsky, and H. Vahlbruch, First Long-Term Application of Squeezed States of Light in a Gravitational-Wave Observatory, Phys. Rev. Lett. 110, 181101.
\bibitem{lab20}Lisa Barsotti, Jan Harms, and Roman Schnabel, Squeezed vacuum states of light for gravitational wave detectors,
\bibitem{lab21}Rep. Prog. Phys. 82 016905 (2019).
U. Dorner, R. Demkowicz-Dobrzanski, B. J. Smith, J. S.
Lundeen, W. Wasilewski, K. Banaszek, and I. A. Walmsley, Phys. Rev. Lett. 102, 040403 (2009).
\bibitem{lab22} R. Demkowicz-Dobrzanski, U. Dorner, B. J. Smith, J. S.
Lundeen, W. Wasilewski, K. Banaszek, and I. A. Walmsley, Phys. Rev. A 80, 013825 (2009).
\bibitem{lab23}F. Hudelist, J. Kong, C. Liu, J. Jing, Z. Y. Ou, and W. Zhang, Nature Communications 5, 3049 (2014).
\bibitem{lab24}Y.-S. Wang, C. Chen, and J.-H. An, New Journal of Physics 19, 113019 (2017).
\bibitem{lab25}Margret Heinze and Robert K$\ddot{o}$nig, Phys. Rev. Lett. 123, 010501 (2019).
\bibitem{lab26}Qi Yao, Jun Zhang, Xiao-Feng Yi, Li You, and Wenxian Zhang, Phys. Rev. Lett. 122, 010408 (2019).
\bibitem{lab27}Lian-Ao Wu and Daniel A. Lidar, Phys. Rev. A 70, 062310
(2004).
\bibitem{lab28}Asoka Biswas and Daniel A. Lidar, Phys. Rev. A 74, 062303 (2006).
\bibitem{lab29}Bibek Pokharel, Namit Anand, Benjamin Fortman, and Daniel A. Lidar, Phys. Rev. Lett. 121, 220502 (2018).
\bibitem{lab30}D. Vitali and P. Tombesi, Phys. Rev. A 59, 4178 (1999).
\bibitem{lab31}D. Vitali and P. Tombesi, Phys. Rev. A 65, 012305 (2001).
\bibitem{lab32}A. W. Chin, S. F. Huelga, and M. B. Plenio, Phys. Rev. Lett. 109, 233601 (2012).
\bibitem{lab33}Dong Xie, An Min Wang, Phys. Lett. A 378, 2079-2084  (2014).
\bibitem{lab34}J. J. Cooper, D. W. Hallwood, J. A. Dunningham, and
J. Brand, Phys. Rev. Lett. 108, 130402 (2012).
\bibitem{lab35}X.-M. Lu, S. Yu, and C. H. Oh, Nature Communications 6, 7282 (2015).
\bibitem{lab36}Yu-Xin Wang and A. A. Clerk, Phys. Rev. A 99, 063834 (2019).
J. Wiersig, Phys. Rev. Lett. 112, 203901 (2014).
\bibitem{lab37} N. Zhang et al., Sci. Rep. 5, 11912 (2015).
\bibitem{lab38}J. Wiersig, Phys. Rev. A 93, 033809 (2016).
\bibitem{lab39}J. Ren et al., Opt. Lett. 42, 1556 (2017).
\bibitem{lab40}W. Chen et al., Nature 548, 192 (2017).
\bibitem{lab41}H. Hodaei et al., Nature 548, 187 (2017).
\bibitem{lab42}T. Ono and H. F. Hofmann, Phys. Rev. A 81, 033819 (2010).
\bibitem{lab43}P. A. Knott, T. J. Proctor, K. Nemoto, J. A. Dunning-ham, and W. J. Munro, Phys. Rev. A 90, 033846 (2014).
\bibitem{lab44}U. Dorner, R. Demkowicz-Dobrzanski, B. J. Smith, J. S. Lundeen, W. Wasilewski, K. Banaszek, and I. A. Walmsley, Phys. Rev. Lett. 102, 040403 (2009).
\bibitem{lab45}A. J. Leggett, S. Chakravarty, A. T. Dorsey, M. P. A. Fisher, A. Garg, and W. Zwerger, Rev. Mod. Phys. 59,
1 (1987).
\bibitem{lab46}V. Weisskopf and E. Wigner, Eur. Phys. J. A 63, 54 (1930).
\bibitem{lab47} W. Louisell, Quantum Statistical Properties of Radiation (Wiley, New York, 1973), pp. 418-428.
\bibitem{lab48}L.-P. Yang, C. Y. Cai, D. Z. Xu, W.-M. Zhang, and C. P. Sun, Phys. Rev. A 87, 012110 (2013).
\bibitem{lab499}H. Wiseman and G. Milburn, quantum measurement and control  Cambridge University Press, New York, 2010).
\bibitem{lab49}S.F. Huelga, C. Macchiavello, T. Pellizzari, A.K. Ekert, M.B. Plenio, J.I. Cirac, Phys. Rev. Lett. 79 (1997) 3865.
\bibitem{lab500}S. Longhi, Opt. Lett. 43, 5371 (2018).
\bibitem{lab501}W. Langbein, Phys. Rev. A 98, 023805 (2018).
\bibitem{lab50}J. Naikoo, K. Thapliyal, A. Pathak, and S. Banerjee, Phys. Rev. A 97, 063840 (2018).
\end{thebibliography}
\end{document}